# HIGH TEMPERATURE SUPERCONDUCTING MAGNETS FOR EFFICIENT LOW ENERGY BEAM TRANSPORT SYSTEMS

J. Nipper, G. Flanagan, R. P. Johnson, Muons, Inc., Batavia, 60510 IL, USA
M. Popovic    Fermilab, Batavia, 60510 IL, USA

*Abstract*

Modern ion accelerators and ion implantation systems need very short, highly versatile, Low Energy Beam Transport (LEBT) systems. The need for reliable and continuous operation requires LEBT designs to be simple and robust. The energy efficiency of available high temperature superconductors (HTS), with efficient and simple cryocooler refrigeration, is an additional attraction. Innovative, compact LEBT systems based on solenoids designed and built with high-temperature superconductor will be developed using computer models and prototyped. The parameters will be chosen to make this type of LEBT useful in a variety of ion accelerators, ion implantation systems, cancer therapy synchrotrons, and research accelerators, including the ORNL SNS. We plan to evaluate the benefits of solenoids made with HTS using analytical and numerical calculations for a single and two-solenoid configuration. The two solenoid study will be tailored toward the system to be used in the SNS prototype LEBT that will replace the electrostatic one at SNS, and a single solenoid configuration, as was proposed for the Fermilab proton driver that will be most applicable to ion implantation applications.

## INTRODUCTION

Low-Energy Beam-Transport (LEBT) systems for ion beams often perform several functions. The main function is to match the emittances of the beam extracted from the ion source into the following accelerator structure, usually a Radio Frequency Quadrupole (RFQ). In order to facilitate a variety of input conditions and allow tuning it is desirable to have a number of parameters to tune the system. This is often achieved by having at least two independent lenses (e.g. solenoids and/or quadrupoles). Additionally, the LEBT may need some beam diagnostics to measure current and emittance. High-performance designs, such as the LEBT for the SNS at ORNL, also require a chopping system to select a portion of the beam to be accelerated. All of these tasks need to be performed while suppressing emittance growth. Emittance growth is a concern especially in the case of high current applications.

The traditional magnetic LEBT for high current ion beams is typically on the order of one meter, or longer. It consists of large aperture magnetic lenses such as solenoids or quadrupole combinations, vacuum system, and some facility for beam diagnostics. However, the desire for beam currents approaching 100 mA has led to a need for a shorter LEBT design to prevent space charge effects from causing excessive emittance growth. This growth with such high current may not be manageable even with beam neutralization using a controlled addition of a compensating gas. In the case of a two-solenoid based LEBT the power consumption can be very high. While such costs may be of minor importance in a large research facility, the can be of major importance in a smaller application, such as ion implantation.

Here we discuss a concept for a compact, highly tuneable, superconducting solenoid LEBT based on the use of the latest HTS in order to facilitate a compact design with a large range of possible fields. The outcome will be a versatile LEBT design that can address the needs of many different applications. Of primary interest is a LEBT that could serve as a replacement to the existing SNS electrostatic LEBT using modern HTS technology with its high efficiency, high-field capability, and excellent radiation tolerance [ref].

## SINGLE SOLENOID HTS LEBT SOLUTION

Here we give an overview of a single solenoid LEBT that was proposed for a superconducting proton linac at FNAL [1], we include this work explicitly as it serves to outline other requirements and a somewhat general scenario for any proposed LEBT.

For a superconducting proton-driver Linac [2], the H-ion beam will be transported and matched from the source to the radio frequency quadrupole (RFQ) using a short low energy LEBT. The present Fermilab ion source produces 45 mA of H- beam at 50 keV, this overview is clearly application-specific, but, could be tuned to other similar applications such as for SNS or ion implanting accelerators.

The design of the transport line is based on the following requirements [3]:

- No electrostatic elements
- No bunching at low energy
- Pressure > 10E-6 Torr to have more than 90% neutralization
- Pressure < 10E-5 Torr to avoid stripping more than 10% per meter
- No potential wall traps for ions
- Fast Ion Source replacement
- Accurate and fast beam diagnostics
- Transport line as short as possible to avoid emittance growth.

The line has one solenoid [4], two trim magnets, two vacuum valves with pumping ports, and a quick disconnect before the second vacuum valve. It should be noted that the primary focus of this proposal is what lies between the trim magnets, the solenoid.

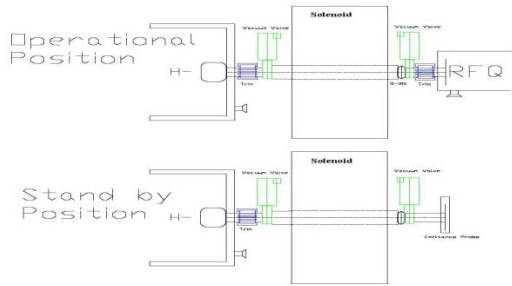

1. Schematic diagram of a single solenoid LEBT solution.

The ion source, which is held at negative high voltage, the grounded column terminal that contains the ion source, the first correction trim magnet package, the first vacuum valve, and the solenoid make up the first unit. The second vacuum valve and the second correction trim magnet package are attached to the RFQ and are a second, separate unit. The main pumping ports are on the grounded ion source container are on the body of the RFQ. The two units are connected together using quick disconnects and the two pumping ports on the gate valves are used to quickly pump down the interconnecting region to operational pressure. The solenoid has an inner bore radius of 50 mm and can be moved along the beam pipe between the first gate valve and the quick disconnect. The total length of beam transport from the source aperture to the matching point of the RFQ is about 700 mm. The distance from the quick disconnect to the RFQ matching point is about 100 mm.

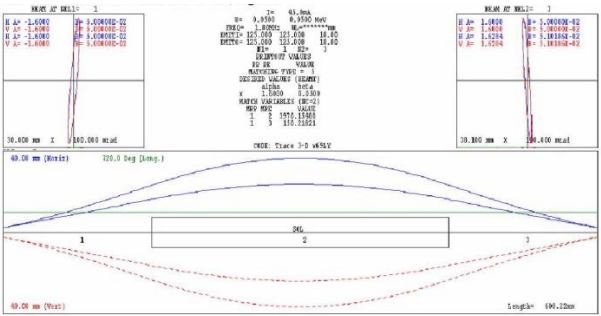

Figure 2. Trace3D solution for the example shown in Fig. 7, showing how the solenoid current can be adjusted to keep the Twiss parameters constant at the entrance to the RFQ for 0 and 40 mA beam current.

Steering errors are corrected using two trim magnets, one right after the source and the other at the entrance to the RFQ. The trim magnets are window frame magnets with horizontal and vertical correction coils on the same frame. The magnet next to the RFQ can provide an angular kick to correct for any positional misalignment of the source or solenoid.

The trim magnet near the source corrects for any misalignment of the source and the ground terminal. This magnet will be used to center the beam in the aperture of the RFQ. The main adjustment will be done in stand-by position using an emittance probe.

We propose to design and build a unit that is very short, easily tuned, and low maintenance. There are two key innovations in this single solenoid design, an HTS solenoid, allowing for a very short LEBT, and the ability to move the solenoid parallel to the beam direction for more tuning capability. The system that houses the solenoid(s) will have the capability to move along the beam line. While the ability to move the cryo vessel is not a consideration for a SNS type replacement, it will help to make the system commercially viable to a wider market. Additionally, due to space limitations the cryo vessel will have to be very compact. To ensure that the system has turnkey operation great care will be given to the design of the cryo, vacuum, and electrical connections. The ability to move the solenoid (either the single coil or the second of a pair) will aid in having a single LEBT design that can handle multi-species running, i.e. the LEBT could be used for $H^+$, $H^-$, or $D^+$ (not at the same time).

## TWO-SOLENOID HTS LEBT SOLUTION

Our concept for an HTS magnetic LEBT with movable cryo-vessel could be designed and optimized for use as a replacement for the existing SNS electrostatic LEBT. The goals of the design are compactness, low power consumption, and large range of operating parameters, or tunability.

Work at the SNS to replace the present electrostatic LEBT with a new two-solenoid magnetic version was recently reported [5]. The normal-conducting solenoid magnets for a prototype have been constructed and simulations to optimize the LEBT are underway [6] under an SBIR-STTR grant with Tech-X. We are interested in determining the positions (position ranges) and the number of superconducting solenoids needed to achieve adequate matching to the input beam parameters of the RFQ. Additionally, the need/specifications for other beam elements will be determined (choppers, trim coils, diagnostics, etc.). If it is found that a single solenoid configuration can provide adequate matching then our design will follow the work outlined in previous description of a single solenoid LEBT. However, the field strengths will now reflect those achievable with a superconducting solenoid. A single solenoid design, using a longer HTS solenoid, may allow for a very compact configuration that is more appropriate for applications like H- injection into proton drivers, ion implantation, or cancer therapy synchrotrons.

## HTS AND DESIGN CONSIDERATIONS

The choice of conductor is not obvious in this application. Both YBCO and Bi2212 have strengths and weaknesses. Bi2212 is the only HTS conductor that can be formed into an isotropic round wire with high $J_c$. This provides the flexibility of making Bi2212 into low aspect ratio rectangular cables or flat tapes. Unfortunately, Bi2212 has exhibited poor electro-mechanical properties and the requirement of a final heat treatment at 890 deg C in pure oxygen limits fabrication to being a wind and react technology.

In comparison to Bi2212, YBCO is fabricated in a continuous deposition process. It is fabricated into a final superconducting product before being sold, and therefore, before magnet fabrication begins. YBCO typically has a nickel or hastelloy substrate, this coupled with the fact that the conducting layer is close to the neutral axis makes for a mechanically strong conductor. The drawback of YBCO is the manufacturing is limited to low aspect ratio tapes (typically 4mm or 12mm wide and 0.1mm thick). The limitation of being available as low aspect ratio tape is not expected to be a concern for this application.

A possible solenoid for the LEBT made using HTS might look as shown in Fig. 4. The red volume is the conductor and the blue represents the iron return yoke. The conductor is 135 mm in length and 40 mm in width. Let us choose SuperPower YBCO SCS120 conductor which is 12 mm wide. The critical current of the doped version of this conductor at 77 K is ~65 amp. This magnet would need 2935 turns of the conductor or 830 m. Fig. 5 shows a contour map of |B| over the solenoid. Figure 6 shows the field on the axis. The engineering current density, $J_E$=30 amp/mm$^2$, corresponding to 54 amp per conductor, was used to calculate the field shown. The peak field in the solenoid is 0.9 T. (If we chose 33 amp, which is half the critical current, a typical safety factor, we would have 0.55 T. This would be compatible with the 0.4 T solenoid used in the single solenoid LEBT model discussed below.)

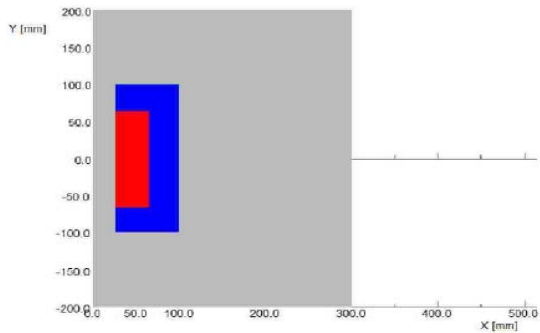

Figure 3. Geometry of an HTS solenoid that has similar field strength and line integral to the ones shown in Fig. 1, but with a radius of only X=10 cm. The length of magnet along the beam direction is Y=20 cm. Both the length of the conductor and the thickness of the iron yoke can be reduced to make a more compact design and/or the strength of the field can be increased to accommodate more advanced LEBT designs.

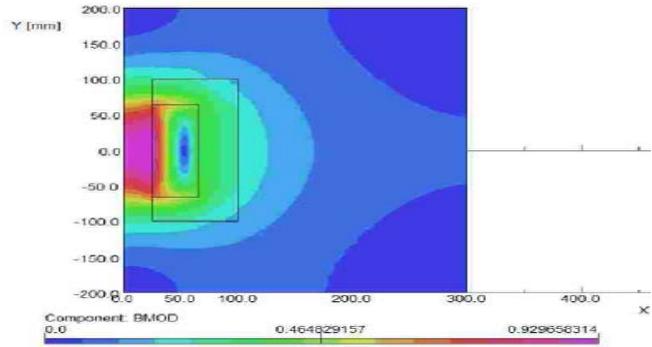

Figure 4. Contour plot of |B| in Tesla.

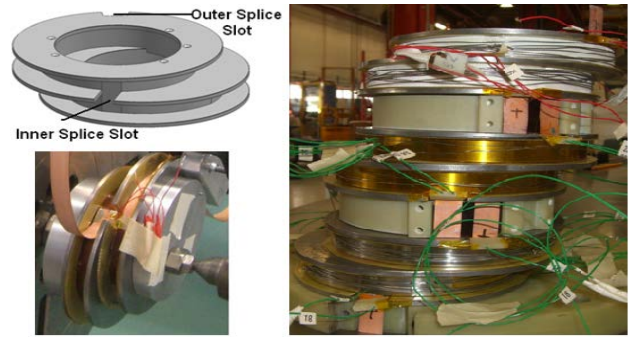

Figure 5. Example of YBCO HTS magnet development that was carried out at the Fermilab Technical Division under an SBIR with Muons, Inc. to develop a high field Helical Solenoid (HS) for 6-dimensional muon cooling. The 12 mm YBCO tape is the same as is proposed for the LEBT application in this proposal. Top Left: the offset spools that make up the double pancake building block for the HS. Lower Left: the winding fixture showing the interface splice joint at the inner layer of the double pancake. Right: Six coils making three double pancake elements are seen, where two dummy RF cavities (seen as gray surfaces with pink covers) separate the double pancakes. These YBCO coils have been tested at 77 K and 4 K and shown to reach about 85% of their short sample limit.